# Multiscale Computing in the Exascale Era

Saad Alowayyed[1,2], Derek Groen[3], Peter V. Coveney[4], Alfons G. Hoekstra[1,5,*]

1   Computational Science Lab, Institute for Informatics, University of Amsterdam, The Netherlands
2   King Abdulaziz City for Science and Technology (KACST), Riyadh, Saudi Arabia
3   Department of Computer Science, Brunel University London, United Kingdom
4   Centre for Computational Science, University College London, United Kingdom
5   ITMO University, Saint Petersburg, Russia
*   corresponding author, a.g.hoekstra@uva.nl

**Abstract**

We expect that multiscale simulations will be one of the main high performance computing workloads in the exascale era. We propose multiscale computing patterns as a generic vehicle to realise load balanced, fault tolerant and energy aware high performance multiscale computing. Multiscale computing patterns should lead to a separation of concerns, whereby application developers can compose multiscale models and execute multiscale simulations, while pattern software realises optimized, fault tolerant and energy aware multiscale computing. We introduce three multiscale computing patterns, present an example of the extreme scaling pattern, and discuss our vision of how this may shape multiscale computing in the exascale era.

## 1   Introduction

Science is at its most powerful when it can not only convincingly explain the processes at work in natural phenomena but is able to predict what will occur before it does so. Predictions of real world events, such as weather forecasting, when a cement will set, the occurrence of an earthquake or what medical intervention to perform in order to save a person's life, all require the bringing together of substantial quantities of data together with the performance of one or more likely many high-fidelity four (three space and one time) dimensional simulations before the event in question occurs. Such forms of calculation are among the most demanding known in computational science, as they need to be done rapidly, accurately, precisely and reliably, including quantification of the uncertainties associated with them. They are also *multiscale* in nature, as their accuracy and reliability depend on the correct representation of processes taking place on several length and time scales. Only now, as we move toward the exascale era in high performance computing (HPC) can we expect to be able to tackle such problems effectively and, eventually, in a routine manner.

Indeed, multiscale phenomena are everywhere around us [1–7]. If we study the origin and evolution of the universe [8] or properties of materials [9–13], if we try to understand health and disease [3,14–21] or develop fusion as a potential energy source of the future [22], in all these cases and many more we find that processes on quite disparate length and time scales interact in strong and non-linear ways. In short, multiscale modelling is ubiquitous and progress in most of these cases is determined by our ability to design and implement multiscale models of the particular systems under study [1,6,23].

The sheer complexity of such multiscale phenomena still limits our capability to perform high-fidelity simulations that accurately and reliably predict the behaviour of a given system in all situations. Capturing even a few of those coupled processes in a multiscale simulation quickly reaches the limits of contemporary high performance computing at the petascale.

That the importance of multiscale modelling in many domains of science and engineering is still increasing is clearly demonstrated in numerous publications; see, for example, [1,24]. Therefore, we must anticipate that multiscale simulations will become an increasingly important form of scientific application on high end computing resources, necessitating the development of sustainable and reusable solutions for such emerging applications, that is, *generic* algorithms for multiscale computing. As we move into the exascale performance era we need to drastically change the way we use HPC for simulation based sciences [25].

For example, on current resources we can simulate composite nanomaterials using tens of millions of atoms, where the interaction potentials rest in turn on electronic structure and atomistic simulations of millions of particles but are performed at higher levels of particulate coarse-graining. Such simulations have already led to ground-breaking insights into chemically specific structural self-assembly and large scale materials property prediction [9]. However, they are still limited to at best volumes of matter whose linear dimensions are on the micrometre scale, well below the size one would need to study, e.g., for the formation of fractures on millimetre scales, let alone to predict performance of materials on macro scales such as arise in typical automotive and aerospace applications. Stepping up to that scale requires not only simulating billions of particles, but also dealing with the non-linear increase in the temporal range that needs to be analysed. This in turn amplifies the need for advanced statistical analysis, which itself becomes a further burden on computational resources. Hence, multiscale challenges such as running atomistic simulations on demand in order to inform a series of coarse-grained simulations and, in future also finite element calculations, will clearly lead us far beyond of what is currently possible at the petascale.

Similar "grand challenges" can be found across the entire scientific spectrum. To get from the current state of the art to more realistic macroscopic regimes requires new developments in multiscale computing to extend discrete representations into the continuum level, optimally designed to exploit an increase in computing capabilities by a factor of between 100 to 1000. Indeed, advanced multiscale algorithms in combination with exascale resources will help us transition to predictive multiscale science. To make this possible, we shall need generic multiscale computing algorithms capable of producing high-fidelity scientific results and scalable to emerging exascale computing systems. We call this *high performance multiscale computing* (HPMC).

In a multiscale simulation, each relevant scale needs its own type of solver. [26,27] Accordingly, we define a multiscale model as a collection of coupled single scale models (loosely defined based on the dominant physical properties that can be computed reliably with a dedicated, so-called "monolithic" solver). We will demonstrate here that one can then identify generic *multiscale computing patterns* (MCPs) arising in multiscale applications that dictate the scope for novel multiscale algorithms at the exascale.

Exascale computing poses a number of key challenges that application developers cannot ignore, such as scheduling and robustness of algorithms and their implementation on millions of processors, data storage and I/O for extreme parallelism, fault tolerance, and reducing energy consumption. [28–31] For these reasons, an incremental approach that attempts to scale up monolithic petascale solutions will not be successful at the exascale. Instead, novel algorithms are needed across the software stack, bridging between the applications and the hardware environments. These algorithms need to be designed specifically to address these exascale challenges in order to guarantee efficiency and resilience. We believe that, drawing on the concept of generic MCPs, we can realise a separation of concerns, where the challenges stated above can be resolved to a large extent on the level of the MCPs, while the multiscale application developers can focus on composing their multiscale simulations. This would then lead to much shorter development cycles for multiscale simulations and much more reliable multiscale computing on exascale machines.

There are also strong computational considerations that dictate a need to shift the paradigm for usage of high performance computers from the conventional promotion of monolithic codes which scale to the full production partition of these computers, to much more flexible computing patterns. This calls for new algorithmic approaches like the ones we introduce here, based on our vision of Multiscale Computing Patterns. To clarify this further, computational scientists have worked out numerous effective ways in which to perform spatial domain decomposition. However, petascale and future exascale machines can only reach these performance levels by aggregating a large number of cores whose individual clock speeds are no longer increasing. As a result these high performance computers are becoming "fatter", not faster and speed-up is only achievable by efficient parallelism over all the cores. But because the parallelism is usually applied to the spatial domain, we are increasingly simulating larger slabs of matter, applying weak scaling by using more particles, a higher grid resolution or more finite elements. Yet it often it is the temporal behaviour that one is really interested in, and that behaviour is not extended by adopting larger computers of this nature, or by making the problem physically larger. Since the scientific problems of interest usually have timescales which scale as a nonlinear function of the volume of the system under investigation, each temporal update requires more wall clock time for larger physical problems. This is in fact a recipe for disaster: *we are not getting closer to studying large space and long time behaviour with monolithic codes*. To be sure, accelerators (such as GPUs) and special purpose architectures [32–34] can speed up many floating point calculations in particular cases such as molecular dynamics, often by a factor between one and ten, but this is not sufficient to bridge the vast timescales of concern that range from femtoseconds to seconds, hours and years; nor indeed to quantify the uncertainty in today's still all to prevalent "one-off" simulations.

What *is* needed are more innovative ways of bridging this divide. Multiscale computing as we propose it, is able to do this by deploying its various single scale component parts across such heterogeneous architectures, mapped so as to produce optimal performance and designed to bridge both time and space scales. Thus, we have embarked upon a programme to efficiently deploy componentised multiscale codes on today's and future high performance computers and, thereby, to establish a new and more effective paradigm for exploiting HPC resources.

The goal of this discussion paper is to share our vision on high performance multiscale computing, introduce some or our initial results, and discuss further research directions and open questions.

## 2   Multiscale Computing

Over the last decade, and with many collaborators, we have developed the so-called Multiscale Modelling and Simulation Framework (MMSF) for designing, programming, implementing and executing multiscale applications [3,26,27,35–42]. This framework has been successfully tested on applications from several fields of science and technology (e.g. fusion [35,43], computational biology [35,44,45], biomedicine [17,35,36,46–52], nanomaterial science [9,13,35], and hydrology [35]). The MMSF offers many benefits: a clear methodology, software and algorithm reuse, the possibility to couple new and legacy codes, heterogeneous distributed computing, and access to unprecedented computing resources [26].

Other approaches, in the same spirit as the MMSF, have been described in the literature. For instance, the U.S. Army Research Laboratory has made important steps forward in establishing multiscale computing approaches for materials modelling, establishing a computational framework for bridging different scales [53,54]. Their primary motivation is to shorten development times and to reduce the cost of evaluating new materials for military use. The framework has been shown to connect between two different scales (macroscopic and microscopic), but can easily be extended to interconnect three or more hierarchical scales. A key feature of their proposed framework is a standalone *Evaluation Module*, which has a light coupling between the macroscopic and microscopic models. This module accepts requests from the macroscopic model to perform runs of the microscopic model, and takes care of the scheduling, execution, and data exchange activities required to fulfil the request. The exchange mechanism is asynchronous, which allows multiple microscopic model runs to be performed and managed concurrently by the Evaluation Module.

We believe that MMSF is a natural starting point to actually reason about generic algorithms for multiscale computing on HPC resources, both in terms of a theoretical approach to multiscale modelling as well as a practical way for multiscale simulations. The MMSF therefore forms our foundation to realise and implement the vision of multiscale computing patterns, leading to an exascale multiscale simulation framework offering generic tools that will help us unleashing the full potential of exascale systems. As we outline below, this requires transforming the current MMSF approach into a new kind of HPC, which we call high performance multiscale computing.

The Multiscale Modelling and Simulation Framework is a theoretical and practical way to model, characterise and simulate multiscale phenomena. We have been developing the MMSF over the past years within the European projects COAST[1] and MAPPER[2]. MMSF currently comprises a 4-stage pipeline, going from developing a multiscale model to executing a multiscale simulation, see Figure 1. We describe MMSF detail in [26,27] and references therein.

First, we model phenomena by identifying relevant processes (which are well described by single scale models) and their relevant scales, as well as their mutual couplings. These couplings that define any multiscale model are clearly application dependent and are not addressed themselves by the MMSF. Although this is the core part of any scientifically meaningful multiscale model (see e.g. [6,7]

---

[1] www.complex-automata.org
[2] www.mapper-project.eu

and references therein), once this has been done, MMSF provides a generic framework for multiscale modelling and simulation [26,35].

The single scale models and their coupling are specified with the Multiscale Modelling Language (MML) [27,40], thereby forming the architecture of a multiscale model. It describes the scales and computational requirements of submodels and any scale bridging components needed.

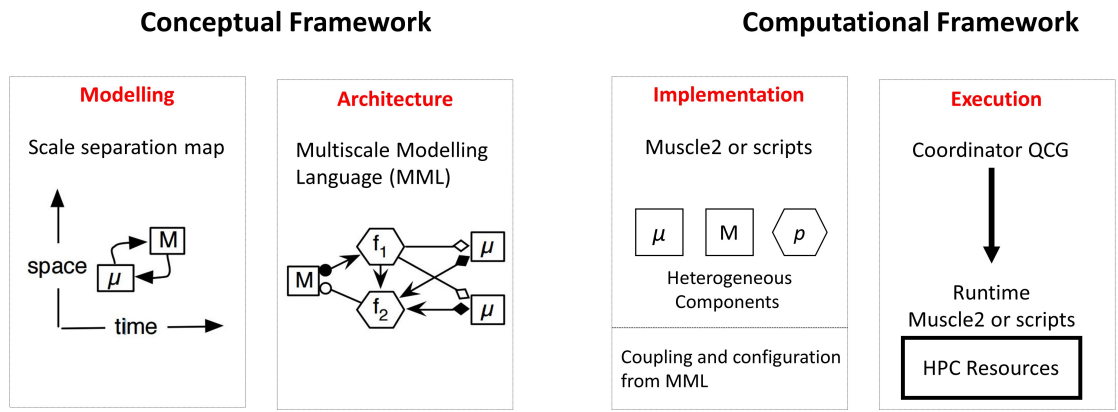

Figure 1: The MMSF pipeline consists of a conceptual part, where the model is specified and coupled (the two panels on the left), and a computational part, where it is implemented and executed (the two panels on the right).

A coupling library like MUSCLE2 [36] ensures that communication between heterogeneous components is possible, with minimal and local changes to the single scale code. In the last step of the pipeline, submodels are executed on suitable computing infrastructure. Each submodel may require different computing resources. Some may be massively parallel and/or may require special hardware and software. In MMSF, the submodels can be distributed on several computers, without additional software development [35,37,47,55].

An important ingredient of MMSF which so far has not been fully exploited is the notion of task graphs for multiscale computing. A task graph is a directed acyclic graph of tasks (the nodes) and their dependencies or data flows (the edges). It can be used for scheduling on parallel and/or distributed computing resources [56]. It can also be seen as a serialized or unfolded graph of the MML description, which may be cyclic. Borgdorff et al. introduced task graphs when specifying the foundations of the MMSF [27], primarily for purposes of deadlock detection, validity checking, and for estimating computational costs and scheduling. As shown in Figure 2, a task graph can be derived from an xMML specification of a multiscale application, which in turn can be used as input for scheduling software. We have demonstrated that task graphs can automatically be derived from xMML [27] and demonstrated the use of task graphs for one specific application [37]. These task graphs will perform an important role in future multiscale applications, since they can be used to create Multiscale Computing Patterns.

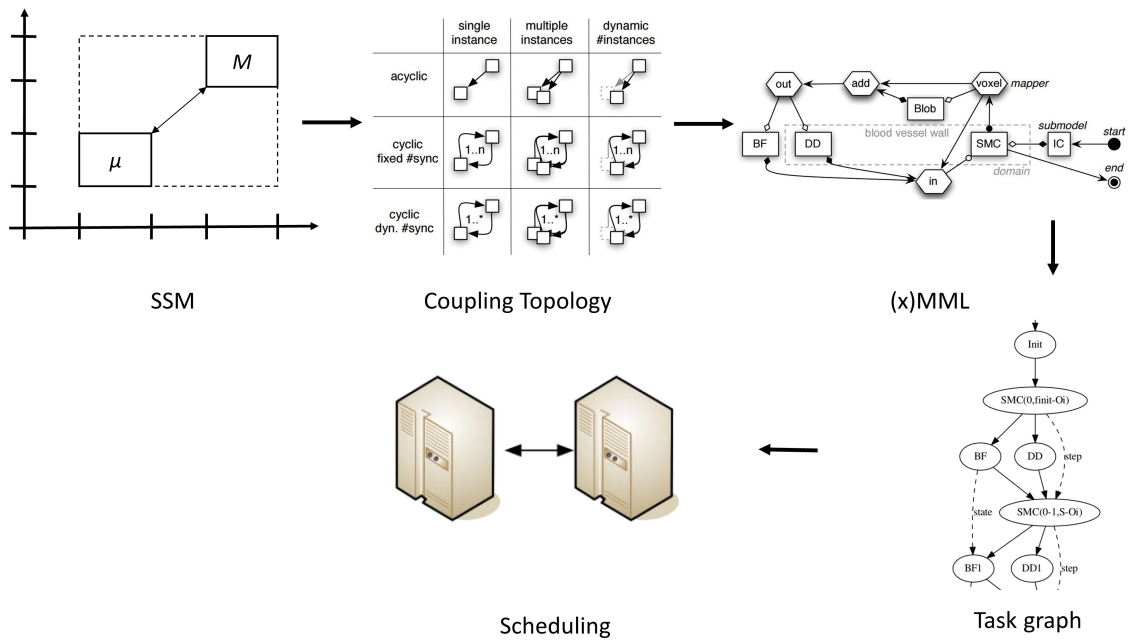

Figure 2: Several stages of description of a multiscale model in the MMSF, starting from the Scale Separation Map, details of the Coupling Topology are added, followed by a full specification in terms of xMML, from which a Task Graph is derived that can then be used as input to scheduling software.

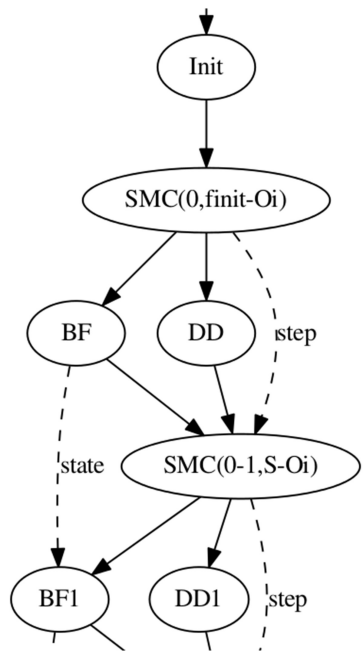

Figure 3: An example of a multiscale task graph, showing the initialisation and first two cycles in ISR3D, an in-stent restenosis model. Note that, in this application when run on a large supercomputer, typically a few thousand complete cycles are performed.

We present an example of a task graph in Figure 3. This relates to a multiscale model for in-stent restenosis, an unwanted response of coronary arteries after balloon angioplasty and stenting. [36,37,51,57,58] The 'SMC' refers to a slow dynamical process, the response of Smooth Muscle Cells in the coronary arterial wall, with a typical timescale of days. Drug Diffusion (DD) simulates diffusion of inhibitory drugs from a stent into the arterial wall and has a timescale of minutes, and the blood flow (BF) has a timescale of one heartbeat, so a second. The growth of SMCs is dictated by their internal biology and at the same time regulated by the blood flow and inhibited by high enough drug concentrations. Given the time scale separation the slow dynamics (SMC) is in quasi-equilibrium with the fast dynamics (DD and BF) that are independent from each other and can be simulated in parallel after each iteration of the slow SMC dynamics. This is clearly expressed by the task graph.

## 3 Multiscale Computing Patterns

Our central assumption is that one can identify generic computing patterns that allow developing algorithms for common scenarios of multiscale computing that cover the vast majority of multiscale scientific applications. We call these multiscale computing patterns. Their main benefit lies in hiding the intricacies of efficient usage of exascale machines, in terms of extreme parallelism and related load balancing, energy awareness and fault tolerance from the multiscale application developers, which should lead to much shorter development cycles and at the same time much more robust and efficient multiscale simulations.

In general, computing patterns can be identified on many levels. They exist deep within the computational kernels of complex application codes (e.g. matrix manipulations relying on BLAS routines, as implemented e.g. in the linear algebra package LAPACK, see, e.g., [59] ) as well as in call sequences involving higher-level subroutines and multiple component codes (e.g. the common component architecture, see e.g. [60]). Since we have already established that there are many generic aspects to multiscale simulations [26,27], exploiting additional genericity is central to the approach we adopt. Indeed, we look for computing patterns at a higher level, namely call sequences of individual single-scale submodels comprising the components of the multiscale model. This leads to only a few multiscale computing patterns.

We define multiscale computing patterns as high-level call sequences that exploit the functional decomposition of multiscale models in terms of single scale models. The task graph introduced in the previous section will be the main ingredient for expressing MCP in terms of objects within the MMSF.

We anticipate a number of key challenges when using today's multi-petascale platforms and future exascale resources in relation to developing and implementing the MCPs. In particular, unprecedented costs in exascale computing have multiple aspects, as running applications on such scales is accompanied with an exceptional cost requirement in terms of computer time (measured e.g. in core or node hours), energy consumption, monetary cost, or time to completion (particularly in urgent computing). Throughout this section we will discuss cost from a generalized perspective, and describe parts in each of the MCPs that are particularly *cost-critical*. Each MCP has different cost-critical parts, and these differences help to more clearly demarcate the definitions for each of the patterns.

Based on our previous experience with multiscale modelling we have identified three computing patterns that we believe are the most pertinent for High Performance Multiscale Computing:

- Extreme Scaling, where one (or a few) single scale models require exascale performance, which are coupled to other, less costly single scale models.
- Heterogeneous Multiscale Computing, where a very large number of microscale models are coupled to a macroscale model.
- Replica Computing, where a large number copies (replicas) are executed, that may or may not exchange information. Replica Computing is plainly in "vanilla" form a single scale pattern, but it can also be regarded as a degenerate form of multiscale computing, in which the models are all at the same scale. There are both incremental variants such as replica-exchange, and the extension to true multiscale modelling which can be based on this form; while the use of replicas at any level of computing provides an important handle on uncertainty quantification.

These three patterns, and combinations thereof, cover a broad range of possible call sequences in high performance multiscale computing. It may turn out that other MCPs should also be considered in due course. These MCPs offer a generic framework for efficient multiscale computing in the emerging exascale performance era, and the three MCPs that we introduce here will prove to be important in that context.

It is important to realise that the MCPs, while being prepared for the challenges listed above, primarily catch the computational structure of the multiscale models, independent of the details of the underlying exascale machines. The latter is hidden in the tools, services and middleware, which the multiscale computing patterns will rely on. This separation of concerns is a main goal of the ComPat project, [61] where we are currently actually implementing the ideas discussed in this paper.

The *extreme scaling* computing (ES) pattern represents a specific class of multi-scale applications where one (or perhaps a few) of the single scale models in the overall multiscale model by far dominates all others, in terms of computational and/or energy cost. Such a dominating *primary model* is expected to scale to very large systems (i.e., multi-petascale or above) and the efficiency of the primary model largely determines the efficiency of the full application. Consequently, one of our goals is to ensure minimal interference by the other single scale models, so-called *auxiliary models*. These typically have a much lower computational and/or energy cost and might even be sequential codes. Load-balancing, decentralized communication, and computation overlapping are some of the techniques we can use here, depending on the relation between the primary and auxiliary models.

The extreme scaling pattern applies when, for instance, in addition to the time and space scale differences, one submodel requires a strong increase in dimensionality or resolution compared to the others, and therefore becomes *cost-critical*. The cost difference between the primary model and the auxiliary models introduces a set of load balancing challenges and possible bottlenecks. The challenge is to efficiently execute the primary model as it is coupled to the auxiliary models, minimizing additional overhead incurred by the coupling and load imbalance, and to propose new variations of the pattern, which include mechanisms for fault-tolerant, energy-efficient, and data-intensive computing.

In the *heterogeneous multiscale computing* (HMC) pattern, we couple a macroscopic model to a large and dynamically varying number of microscopic models. The pattern is based on the heterogeneous multiscale method (HMM) [2,62], which are a class of modelling approaches wherein the constitutive equations of a local state in the macroscale system of interest are not known in closed form, mainly as a consequence of the complexity of the processes at the microscale. The basic philosophy of HMM is to apply a numerical solver to the macroscale equations and to provide the missing macroscale data using an appropriate microscale model. HMMs hold the promise to simulate very complex phenomena, directly coupling detailed microscopic dynamics to emerging macroscopic behaviour. Instead of relying on heuristic constitutive relations or on idealised theory containing simplifying assumptions, HMM allow all microscopic details to be taken into account while at the same time being able to simulate macroscale dynamics. As examples we can mention HMM models for suspension flow [42] or for galaxy merger simulations [63]. HMM-type high performance computational frameworks are still rare; we seek to establish one by defining a multiscale computing pattern for HMM, which is HMC.

The number of microscale models required in HMC depends on spatial properties of the macroscale model, and can in some cases easily be in the order of $10^6$ or more (e.g., for the suspension flow, where a microscale simulation is required for each lattice point in the macroscale simulation). In addition, each microscale model can be a detailed 3D simulation requiring substantial parallel computing resources (e.g. a cell based blood flow model, explicitly taking individual red blood cells into account [64]) The large number and size of the microscale models causes them to dominate the computational and energy cost of the application as a whole, and they are therefore cost-critical. Consequently, it is essential that these models are efficiently mapped and executed on (exascale) HPC resources, and that any overheads incurred by interactions with other components are minimized. Additionally, the cost of microscale models in HMC, especially when mapped to exascale resources, may result in unprecedented data challenges, which can be addressed by introducing a scalable data management software architecture.

As part of this architecture, an on-the-fly database is needed to limit the number of required microscale simulations [65,66]. This database serves to store previously computed data and, where desirable, interpolate between already computed values to provide input to the macroscale model. This is feasible because the amount of data passed up to the macroscale model is usually not large, perhaps a few floating-point numbers representing quantities of interest (such as the viscosity for fluid problems) for each microscale model.

The on-the-fly database will be managed by an HMC manager. Here the coarse-scale model first sends a request to an HMC manager for the properties that it requires more information. The manager then consults the database for the needed information. A user-defined algorithm will decide whether the cached information is sufficiently accurate, and whether interpolated values may be used. If the available information is insufficient, the manager will start a new microscale simulation to get more accurate fine-scale information. For simulations where the evaluation of cached results is compute intensive, this evaluation will be delegated to a separate computing resource in order not to overload the HMC manager. By using asynchronous I/O and offloading intensive calculations, the HMC manager is able to handle many requests simultaneously.

*Replica Computing* (RC) is a multiscale computing pattern that combines a potentially very large number of terascale and petascale simulations (also known as 'replicas') to produce scientifically important and statistically robust outcomes. The replicas are not part of a larger spatial structure (as is the case in, for example, HMC), but they are applied to explore a system under a broad range of conditions. A good example of Replica Computing is the Binding Affinity Calculator, a quite involved workflow in which production scale molecular dynamics is performed to compute binding affinities between small compounds and proteins, typically followed by similar compute-intensive processing of the trajectory data accumulated in that stage [67–69]. Another example is daily weather forecasting, where many replicas are executed in order to asses uncertainties in the predictions [70,71]. Replica Computing is set up through an initialization stage, which determines the simulations required to explore or incorporate a given parameter space. This initialisation is then followed by one or more sequences of simulation and data processing. In general, within Replica Computing we distinguish three scenarios:

1. *Ensemble simulations*. Here a large set of models (replicas) is initiated, run and analysed. The results of these models, which can either be large scale computing jobs themselves or require small amounts of computing resources, are then provided as part of the initial conditions of one or more secondary models. These models may operate on a much larger space and time scale and in that case are usually critical to the overall application in terms of computational and energy cost, or may need just a fraction of resources as compared to the replica computations.
2. *Dynamic ensemble simulations*. Similar to ensemble simulations, here the approach relies on a set of small-scale models that are initiated, run, and analysed. However, in dynamic ensemble simulations, the analysed data is used to rerun a new, better calibrated or complementary set of small-scale models, allowing characterizing system behaviour in complicated parameter landscapes. This landscape in itself can be multiscale. The results of these model executions are then used as a basis for one or more secondary models, similar to the ensemble simulation scenario.
3. *Replica-exchange simulations*. This scenario introduces a set of models with varying parameters (e.g., temperature or spatial characteristics), which are run concurrently. Simulation data is exchanged between these single-scale models at runtime, for example to allow individual particles to migrate from one model to another. These exchanging replicas may be multiscale in their own right, or they can provide statistically robust data, which is then used for more coarse-grained models, which operate on larger time and length scales. Replica exchange simulations are already used in a variety of fields on smaller scales, including e.g. materials science [72], climate sciences [73], biomedicine [74], and origin of life studies [75].

## 4   Computational Challenges at the Exascale

As exascale architectures are being designed and developed, it has become obvious that these new architectures will introduce a range of new computational challenges. Today's largest (and 3$^{rd}$ most efficient) supercomputer, TaihuLight relies on low-power compute units at extreme parallelism (~11 million cores), accompanied with limited memory capacity (0.1 GB/core, whereas >1 GB/core used to be commonplace before). In general, the TOP500 list shows a clear trend towards heterogeneous systems, and we expect that exascale systems will present us with significant non-uniformity of performance and reliability. Here we elaborate on key exascale challenges and describe how MCPs can help to address them efficiently. Our objective is to address all of these exascale challenges as far as possible on the generic level of the MCPs, as it will allow us to formulate unified solutions for these challenges across a class of applications.

The *extreme parallelism* presented by the massive core counts implies a need for advanced load balancing and scheduling to maximally exploit the offered compute power. The use of MCPs instead of monolithic applications already helps to accommodate this, but we additionally need to develop scheduling and load-balancing algorithms that account for the requirements of each respective MCP in terms of compute efficiency, memory efficiency, and energy efficiency. Constrained optimisation approaches can help us to introduce pattern-level optimizations, incorporating awareness of other algorithmic components of the multiscale simulation framework and identifying common scheduling and load-balancing trade-offs for MCPs.

Failure rates tend to scale linearly with the number of cores, and are expected to be commonplace for models using extreme parallelism. We therefore require *fault tolerant* software algorithms, ideally defined on the generic level of MCPs, that are capable of handling hardware failures by applying heuristics for fault recovery, need for recomputation, or accounting for missing or incomplete data (e.g., using statistical techniques such as data imputation or maximum likelihood). The heterogeneity of the architecture may also lead to non-uniform reliability, and scheduling algorithms could support deploying the cost-critical components in our MCPs on resources with a higher reliability.

A main constraint for exascale systems will be *power consumption*. We therefore need to take energy efficiency into account throughout the whole development process of the multiscale computing algorithms, and deploy multiscale computing applications such that we achieve the best trade-off between performance and power consumption. [76] To accomplish this, we should introduce energy efficiency as a constraint parameter in global optimisation algorithms, and identify regimes where MCPs can take advantage of hardware capabilities such as reducing clock frequency or switching cores on/off dynamically.

## 5   Generic Task Graphs and Multiscale Computing Patterns

An important next step, as we will argue, is the realisation that the MCPs can, in some form, be expressed on the level of the task graph. The task graph, as explained above, is a directed acyclic graph used to determine the execution order of submodels, to schedule submodel dependencies, and to estimate runtime and communication cost. The key idea is that we define generic task graphs for each MCP, such that application specific task graphs can be embedded in the generic task graphs. This embedding should be automated. Next, we use the generic task graph to obtain an optimized mapping of the application to an HPC resource, and try to find generic algorithms for this. What exactly is meant by an 'optimal' mapping needs to be defined, or can be made application specific. In any case, it should be optimized with respect to several dimensions (efficient use of resources, power consumption, wall clock time, load balancing, fault tolerance). The way to proceed is that for each generic task graph we need to specify sets of optimal execution profiles, or define constrained optimization problems that should be easily solvable when fed with details of the specific applications. Figure 4 summarizes the approach. An MCP is a tuple of a generic task graph plus data or models on the performance of single scale models (left in Figure 4), a specification of a specific multiscale application in terms of the MMSF (right in Figure 4) and a set of algorithms and heuristics that combine this into detailed input/configuration files for the execution environment in which the multiscale simulation will be executed.

To summarize, the approach that we follow will be that a task graph is generated from the application specific xMML description. This is then taken together with execution recipes specific for an MCP and performance models, or data, for the single scale models, to determine the actual execution profile. This will finally be specified in configuration files for the execution middleware. This immediately leads to a number of key questions:

1. Can we find such generic task graphs for each pattern?
2. How can we map specific applications to such task graphs?
3. How can we use the generic task graphs to set up optimized executions on HPC machines?
4. What information is needed and on which level to make this work?

In the remainder of this paper we formulate some partial answers and work out one example. In future work we intend to generalise these findings to create full-fledged MCPs.

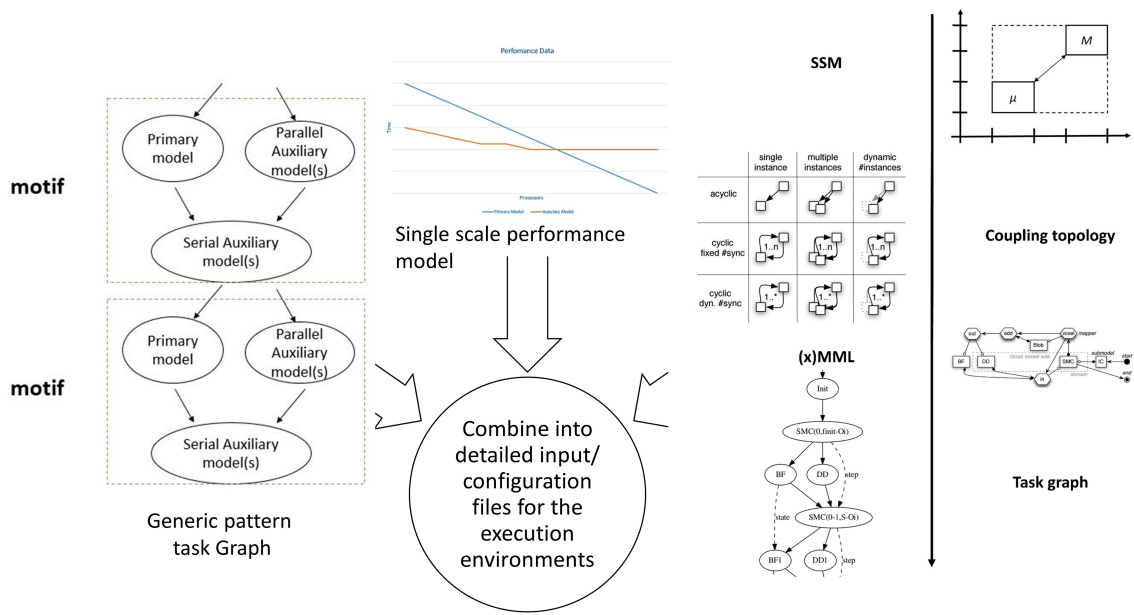

Figure 4: Multiscale computing patterns implemented as generic task graphs and algorithms to generate sufficient information for the execution engines.

We have worked out generic task graphs for the ES, the HMC, and RC patterns, see Figure 5 to Figure 7. The meaning of the boxes and symbols in these figures is shown in the legend in Figure 8.

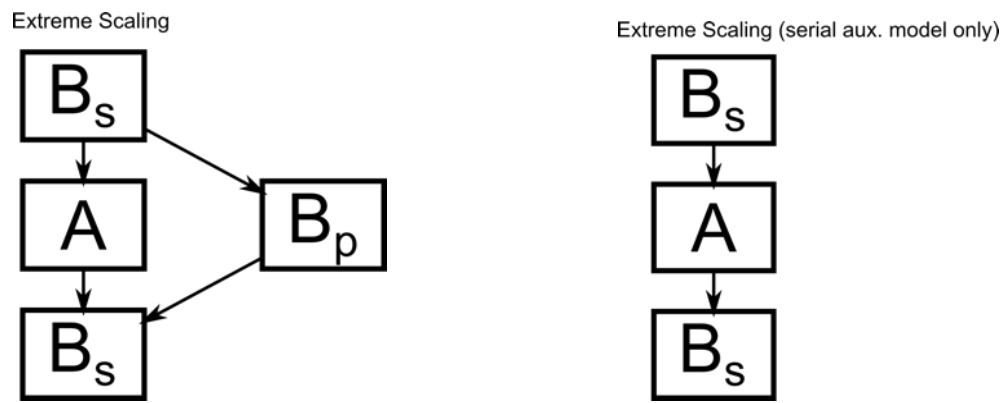

Figure 5: Generic task graph for ES. The version on the left shows the graph with both serial and parallel auxiliary models, the version on the right only has a serial auxiliary model.

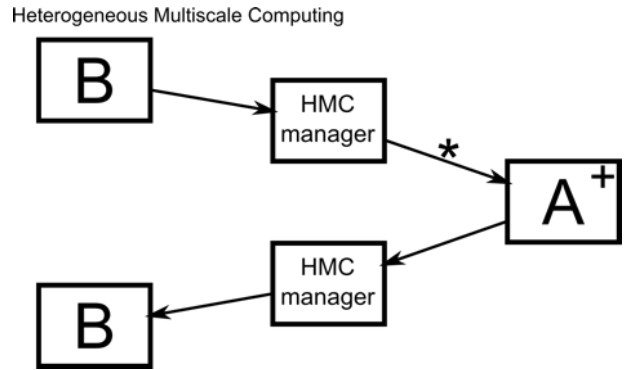

Figure 6: Generic task graphs for HMC.

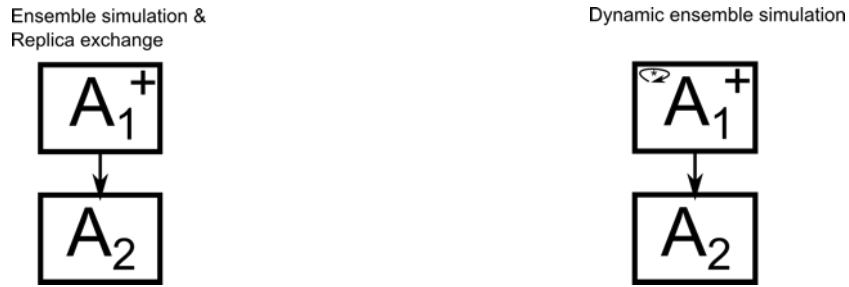

Figure 7: Generic task graphs for RC. The version on the left shows the static variant for ensemble simulations and replica exchange, while the version on the right is for dynamic ensemble simulation.

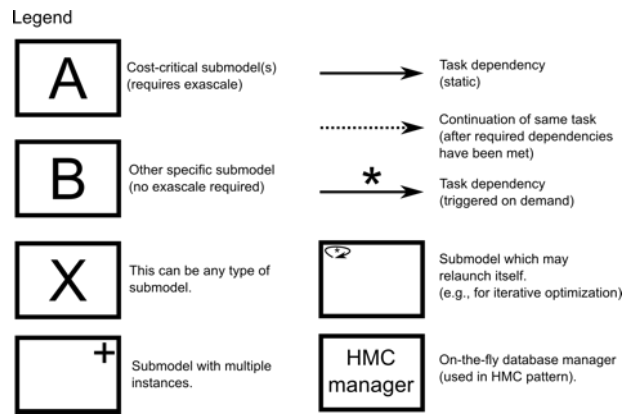

Figure 8: Legend used in the generic task graphs.

Figure 5 (left) shows the generic task graph for ES, where a collection of auxiliary models can either be executed in parallel with the primary model, or in series with the primary model. Figure 5 (right) shows the version where only a serial auxiliary model is present. The graphs should be understood that they are repeating elements in the overall task graph. So, considering the task graph for the In-Stent Restenosis application in Figure 3, and knowing that the fluid flow process is cost-critical, we can identify the BF (blood flow) process with A in the generic task graph, the DD (drug diffusion) process with the parallel auxiliary model $B_p$ in the generic task graph, and SMC (Smooth Muscle Cells) iterations together as the serial auxiliary model $B_s$ in the generic task graph. We find many repeating units of the generic task graph for ES in the task graph from Figure 3, typically a few 1000. We also find that the application in Figure 3 has an initialisation phase as well as a post processing phase (not shown in Figure 3), which are currently not captured by the generic task graphs.

Depending on the execution behaviour of the primary and auxiliary models on HPC machines, a specific execution of the ES graph is considered. The main aim is to align the auxiliary models to produce their data before the primary model requires it. This would be achieved by either pre-computing the values of non-scaling models or interleaving simulations as suggested by the performance model, see also the discussion in the next section.

For HMC, Figure 6, a large and dynamic number of microscale simulations is coupled with one macroscale model, with a HMM database in between. The role of the database is to prevent computing of previously computed results, to interpolate between earlier computed results, and to submit microscale simulation jobs when needed. The latter could be done in a pro-active way, if resources allow it, to start precomputing quantities in anticipation of request from the macroscale solver. A specific execution graph for this pattern depends on the execution behaviour of the macro- and micro-scale models on HPC machines.

For RC, see Figure 7, we find two variants, which capture the behaviour of the three types of replica computing that we defined. In both cases a potentially large set of replicas $A_1^+$ are executed independently and then feed into a second master process $A_2$. In RC, if a replica fails, a restart is not immediately needed, as long as the overall statistical quality of the ensemble that is computed by the RC application is maintained. This is a main distinction with HMM, where if a microscale simulation fails it must be restarted, as the database requested output from this microscale simulation.

# 6 Example of an extreme scaling pattern

A very common example of a multiscale model is a multi-domain scenario [27] where in one part of a computational domain a microscale simulation is coupled to a macroscale simulation in another part of the computational domain. This is typical in situations where one needs to zoom into molecular or even quantum-mechanical details in small regions of space, as e.g. in the classical example of multiscale modelling of crack propagation in materials [77]. Although the microscale simulation will usually take up only a very small portion of the computational domain, the required computational resources for the microscale simulations are orders of magnitude larger than for the macroscale simulation. This immediately puts this scenario into the Extreme Scaling MCP, where the microscale simulation is the primary model, and the macroscale simulation the auxiliary model. The application specific scale bridging dictates if the macroscale simulations are serial or parallel auxiliary models. Regardless of these details, an important question is how the available computational resources should be distributed over the microscale and macroscale simulations (as already shortly discussed in the pioneering work by Broughton et al [77]), how fault tolerance should be realised, and how such simulations could be made energy aware. A first example of how this may be realised on the level of the MCP is discussed below.

One of our own applications is in the field of computational biomedicine [47]. We are in the process of developing the "Virtual Artery" [48], where cell-based (microscale) models are coupled to continuous tissue (macroscale) models in a multi domain scenario. As an example, we consider the case of a cell-based blood suspension model as the primary model, coupled to continuous blood flow models as the auxiliary models, see Figure 9. The idea would be that in an artery, where blood flow is modelled by a continuous flow solver, there are small regions where we actually want to model the behaviour of the red blood cells and platelets. This could be in an aneurysm, for example, to better understand possible thrombosis inside the aneurysm cavity [78].

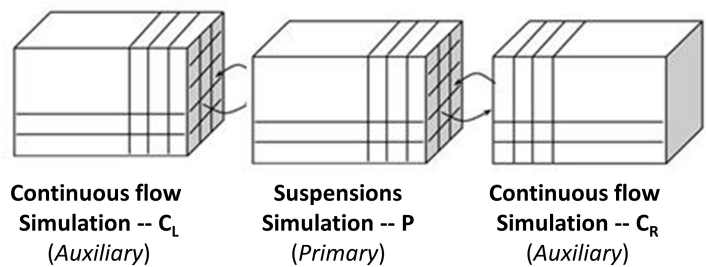

**Continuous flow Simulation -- $C_L$** (*Auxiliary*)   **Suspensions Simulation -- P** (*Primary*)   **Continuous flow Simulation -- $C_R$** (*Auxiliary*)

Figure 9: Example of an ES application, a suspension simulation (the primary model) coupled to two continuous blood flow models (auxiliary models) that provide the in- and outflow conditions for the suspension model.

The resulting Scale Separation Map, gMML and task graph are shown in Figure 10. Note that we assume two instantiations of the auxiliary model representing the inlet and outlet regions of the suspension flow domain. Also note that the embedding of this application task graph into the generic ES task graph (Figure 5, left) is straightforward, the two instantiations of continuous flow solver together form the serial auxiliary model, and the parallel auxiliary model is empty.

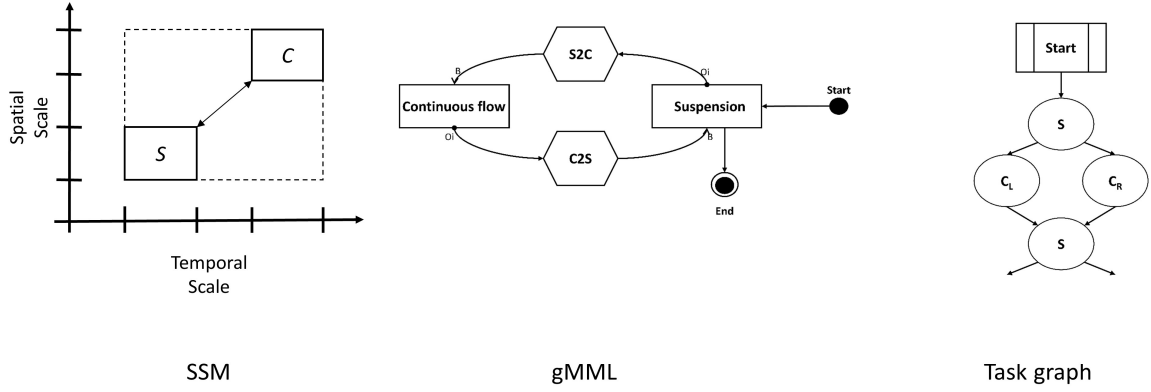

Figure 10: The SSM (left), gMML (middle), and task graph (right) for the extreme scaling multiscale computing pattern coupling a primary model for blood suspensions to two instantiations of the auxiliary model.

We formulate a generic and straightforward performance model for this ES MCP. We assume that we have all information on the performance of the single scale models that make up the multiscale application. This is in line with the MMSF philosophy, the single scale models and the scale bridging algorithms are available and implemented on available resources. The MMSF provides the means to compose the full multiscale model and execute it in an efficient way on available resources.

Call $T_{pr}(P, N_{pr})$ the execution time of the primary model as a function of the number of processors $P$ and the problem size of the primary model $N_{pr}$. Likewise, call $T_{aux}(P, N_{aux})$ the execution time for the serial auxiliary model as of function of $P$ and the problem size of the auxiliary model $N_{aux}$. A defining feature of the ES patterns is that the primary model is very compute intensive and needs petascale or even exascale performance. In terms of the performance model this means that $T_{pr}(1, N_{pr}) \gg T_{aux}(1, N_{aux})$. Both the primary and auxiliary models can run in parallel, but their scalability can be completely different. For the execution time of the ES scenario we can now write

$$T_{ES} = T_{pr}(P, N_{pr}) + T_{aux}(P, N_{aux}) \qquad (1)$$

and for the resulting efficiency of the ES pattern we can write

$$\varepsilon_{ES} = \frac{1}{P}\left(\frac{T_{aux}(1)+T_{pr}(1)}{T_{aux}(P)+T_{pr}(P)}\right) \approx \frac{1}{P}\left(\frac{T_{pr}(1)}{T_{aux}(P)+T_{pr}(P)}\right) = \frac{1}{P}\left(\frac{1}{\frac{T_{aux}(P)}{T_{pr}(1)}+\frac{T_{pr}(P)}{T_{pr}(1)}}\right) = \left(\frac{\varepsilon_{pr}}{\frac{T_{aux}(P)}{T_{pr}(P)}+1}\right) \qquad (2)$$

with $\varepsilon_{pr}$ the efficiency of the primary model, and where for convenience we have dropped the dependence on the problem size. We assume that the primary model scales very well, which seems like a reasonable assumption, given that in the ES case the primary model is the one that requires almost all computing resources and has been optimized sufficiently to run very efficiently on Petascale or emerging Exascale resources.

We consider two limiting cases. First assume that the auxiliary model does also scale well enough on the available resources, at least so that even on $P$ processors, $T_{pr}(P, N_{pr}) \gg T_{aux}(P, N_{aux})$. In this case we find that $\varepsilon_{ES} = \varepsilon_{PR}$ and we can simply execute the ES task graph, as drawn in Figure 11(left).

The other extreme would be that the auxiliary model does not scale at all, even to the point that $T_{pr}(P, N_{pr}) \ll T_{aux}(P, N_{aux})$, leading to $\varepsilon_{ES} \to 0$. This requires a completely different execution of the MCP. Suppose that we could find a pair of processor numbers such that $P = P_1 + P_2$ and $T_{pr}(P_1, N_{pr}) \sim T_{aux}(P_2, N_{aux})$, then we could interleave two instantiations of the ES patterns, combining the execution of a primary model in the first instance with execution of the auxiliary model in the second instance, see Figure 11 (right).

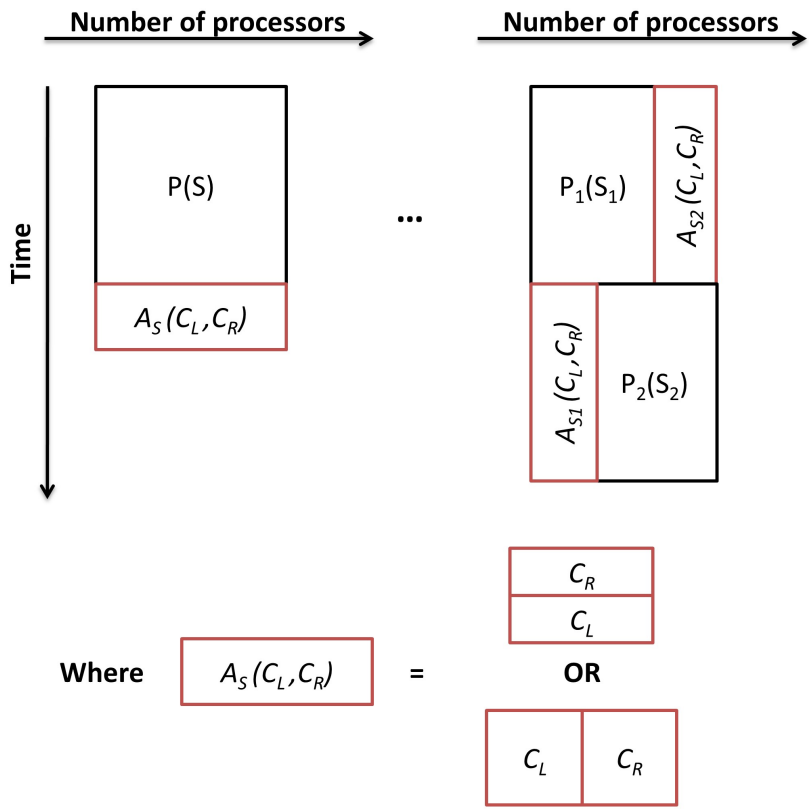

Figure 11: Two limiting possibilities of executing the ES task graph from Figure 10. In the left figure, F denotes Ficsion (the primary model) and $P_1$ and $P_2$ the two serial auxiliary models (Palabos). In the right

figure $F_1$ and $F_2$ denote two instantiations of the primary model, and now $P_1$ and $P_2$ denote the auxiliary models coupled to $F_1$ and $F_2$ respectively (so lumping together the two Palabos instantiations).

In reality most cases will be somewhere between these two limiting cases. Here interesting possibilities appear to optimize both throughput of an ES pattern, to have a well-balanced load, and to optimize energy usage, by allowing adjustment of the clock speeds of the set of processors $P_1$ or $P_2$. In that case $T_{pr}(P_1, N_{pr}) \sim T_{aux}(P_2, N_{aux})$ leads to a whole series of potential solutions of the load balancing problem, and on that manifold we can then optimize either throughput, and/or energy usage of the execution. The ES MCP algorithm will also take into consideration threshold values, which indicate when to switch from one type of execution to another. It is clear that including a parallel auxiliary model into the picture results in an even richer set of potential solutions, and it is the task of the MCP to find optimal execution scenarios, where 'optimal' can be defined in many ways.

## 7 Discussion and Conclusions

We define multiscale computing as the orchestrated execution of coupled single scale models and scale bridging methods that together compose a multiscale model. In our experience, corroborated by practice (see e.g. [35,47]), such componentisation of multiscale computing (as opposed to writing monolithic codes for multiscale models) results in the flexible and efficient development and execution of multiscale simulations. In our opinion, this is the only sensible way to fully exploit all the intricacies of state-of-the-art high end computing resources for multiscale simulations. We have already demonstrated the benefits in distributed computing environments (distributed multiscale computing) [35–37], including clouds [79].

In this paper we have proposed how multiscale computing can be enhanced to fully exploit current petascale and emerging exascale HPC systems. Whereas the componentisation of multiscale computing has already resulted in a very efficient 'lego-based' philosophy for developing multiscale simulations, as well as the benefit of relative straightforward distributed execution, the challenges of computing on high-end HPC machines require additional functionalities in relation to load balancing, fault tolerance, and energy awareness.

In our vision, on the level of multiscale computing, these exascale challenges can be addressed in a generic way, leading to a separation of concerns. The multiscale application developer can keep relying on the lego-based approach, keeping in mind that implementations of the single scale models on HPC resources should already be of very high quality. In many cases the single scale models are readily available, as highly efficient public domain or proprietary codes developed over many years or decades by substantial communities. The execution environment, as offered for instance by the MMSF, should then takes care of the load balancing, fault tolerance and energy awareness of the overall multiscale simulation.

The notion of a small number of generic multiscale computing patterns that capture repeating motifs in multiscale computing task graphs facilitates this separation of concerns. Given sufficient information on the performance and energy profiles of the single scale models, it will be possible to develop MCP software that takes as input the xMML description of the multiscale model and the performance and energy profiles of the single scale models, and generates execution scripts that a middleware can then use to optimally allocate resources and execute the multiscale simulation. The example that we presented in section 6 sheds some light on how we expect that MCPs can be further developed. A large consortium composed of European Universities, HPC centres and companies are currently implementing these ideas as part of the ComPat project [61].

Once available, and if successful, the MCPs, in combination with the already existing MMSF, have the potential to dramatically increase the effectiveness with which we can develop, deploy and execute multiscale simulations on emerging exascale resources.

# 8 Acknowledgements

We acknowledge partial funding from the European Union Horizon 2020 research and innovation programme for the ComPat project (http://www.compat-project.eu/) under grant agreement no. 671564, and the CompBioMed project (http://www.compbiomed.eu/) under grant agreement 675451. S.A. acknowledges funding by King Abdulaziz City for Science and Technology (KACST), Saudi Arabia; A.G.H. acknowledges partial financial support by the Russian Scientific Foundation, grant no. 14-11-00826; and P.V.C, thanks the UK Medical Research Council for a Medical Bioinformatics Grant (MR/L016311/1), along with special funding from the UCL President & Provost.